\documentclass{iopconfser}
\pdfoutput=1
\usepackage[T1]{fontenc}
\usepackage{xspace}
\usepackage{todonotes}
\usepackage{hyperref}
\usepackage{amsmath}
\providecommand{\PH}{\ensuremath{H}\xspace} % plain Higgs
\providecommand{\PZ}{\ensuremath{Z}\xspace} % Z
\providecommand{\PW}{\ensuremath{W}\xspace} % W
\providecommand{\Pg}{\ensuremath{g}\xspace} % g
\providecommand{\PQc}{\ensuremath{c}\xspace} % c
\providecommand{\PAQc}{\ensuremath{\overline{c}}\xspace} % cbar
\providecommand{\PQb}{\ensuremath{b}\xspace} % b
\providecommand{\PAQb}{\ensuremath{\overline{b}}\xspace} % bbar
\providecommand{\PQt}{\ensuremath{t}\xspace} % t
\providecommand{\PQq}{\ensuremath{q}\xspace} % q
\providecommand{\PAQq}{\ensuremath{\overline{q}}\xspace} % qbar
\newcommand{\bbbar}{\ensuremath{\PQb\PAQb}\xspace}

\newcommand{\ccbar}{\ensuremath{\PQc\PAQc}\xspace}
\newcommand{\qqbar}{\ensuremath{\PQq\PAQq}\xspace}
\newcommand{\pt}{\ensuremath{p_{\mathrm{T}}}\xspace}

\newcommand{\ptvecmiss}{\ensuremath{{\vec p}_{\mathrm{T}}^{\kern1pt\text{miss}}}\xspace}

\begin{document}

% \title{Leveraging Large-Scale Pretraining for Efficient Jet Classification: An Evaluation of Transfer Learning and Dataset Scaling in Particle Physics}
\title{Large-Scale Pretraining and Finetuning for Efficient Jet Classification in Particle Physics}

\author{Zihan Zhao$^{1}$, Farouk Mokhtar$^{1}$, Raghav Kansal$^{1}$, Haoyang (Billy) Li$^{1}$, and Javier Duarte$^{1}$}

\affil{$^1$University of California San Diego, La Jolla, CA 92093, USA}

\email{ziz078@ucsd.edu, fmokhtar@ucsd.edu, rkansal@ucsd.edu, hal113@ucsd.edu, jduarte@ucsd.edu}

\begin{abstract}
This study introduces an innovative approach to analyzing unlabeled data in high-energy physics (HEP) through the application of self-supervised learning (SSL).
Faced with the increasing computational cost of producing high-quality labeled simulation samples at the CERN LHC, we propose leveraging large volumes of unlabeled data to overcome the limitations of supervised learning methods, which heavily rely on detailed labeled simulations. By pretraining models on these vast, mostly untapped datasets, we aim to learn generic representations that can be finetuned with smaller quantities of labeled data. Our methodology employs contrastive learning with augmentations on jet datasets to teach the model to recognize common representations of jets, addressing the unique challenges of LHC physics.
Building on the groundwork laid by previous studies, our work demonstrates the critical ability of SSL to utilize large-scale unlabeled data effectively.
We showcase the scalability and effectiveness of our models by gradually increasing the size of the pretraining dataset and assessing the resultant performance enhancements.
Our results, obtained from experiments on two datasets---JetClass, representing unlabeled data, and Top Tagging, serving as labeled simulation data---show significant improvements in data efficiency, computational efficiency, and overall performance. These findings suggest that SSL can greatly enhance the adaptability of ML models to the HEP domain. 
This work opens new avenues for the use of unlabeled data in HEP and contributes to a better understanding the potential of SSL for scientific discovery.
\end{abstract}

\section{Introduction}
\label{sec:intro}
To enable precision measurements of the standard model (SM) and searches for new physics at the CERN LHC, physicists train machine learning (ML) models using detailed, labeled simulations of proton-proton collisions for a variety of tasks including triggering~\cite{Duarte:2018ite,CMSP2L1T}, charged particle tracking, calorimetry~\cite{Bhattacharya:2022sni}, particle-flow reconstruction~\cite{Pata:2022wam}, and jet tagging~\cite{Moreno:2019neq,Qu:2019gqs,Qu2022} and mass regression.
These trained ML models are subsequently applied to real data.
This paradigm is called \emph{supervised learning} because it uses explicit labels derived from simulation settings, e.g., whether a signal or background event is simulated.

However, a significant issue is ensuring that the performance of these ML models trained in simulation translates to real data, especially when the conditions of the two settings are different.
For example, data often has a larger number of average simultaneous proton-proton interactions, known as \emph{pileup} (PU), run-dependent detector noise, and different trigger efficiencies than simulation.
This general problem of different conditions between the training setting and the inference setting has been studied in ML and particle physics and is known as \emph{domain adaption}~\cite{Clavijo:2020mua,Baalouch:2019fhm}.
Typically, these differences are corrected by applying data-simulation scale factors parameterized as a function of a small number of features.
Nonetheless, in recent searches for high-momentum Higgs boson production~\cite{CMS:2022dwd,CMS:2022nmn}, this scale factor measurement and application can be a dominant source of systematic uncertainty.

In this paper, we propose to apply a generalized ML approach, in which models are first pretrained on large quantities of unlabeled data and subsequently \emph{adapted} or \emph{finetuned} using smaller quantities of labeled data for a specific downstream task.
In the pretraining stage, models are trained to learn generic representations of the input features.
% Based on this pretraining, these models learn about the complex relationships among the real data via abstract representations and apply it to reconstruct Higgs bosons and other particle physics tasks.
These representations can then be used to train a smaller, simpler model in smaller quantities of simulation.
Generally, pretraining is achieved through self-supervised learning (SSL), in which portions of input data are masked, paired together, or augmented and the model is tasked to reconstruct the missing data, find matching pairs, or identify data augmentations, respectively.
This forces the model to learn the context of and correlations among elements in the data.
The specific approach we take in this paper is contrastive learning, in which multiple views of the same data are input to the model, and the model learns common representation.
In our case, the algorithm's inputs are the original jets and their augmented versions.
% While the dataset required for pretraining may be massive, it does not need task-specific labels, which are labor-intensive to generate and required for supervised learning.

There are two primary objectives of this approach:
\begin{enumerate}
    \item To demonstrate gains in performance in the smaller labeled dataset by pretraining on larger unlabeled dataset.
    \item To demonstrate the effect of scaling up the sizes of the large unlabeled dataset on the performance of SSL models.
\end{enumerate}

By applying this strategy to particle physics, we can simultaneously (1) leverage copious amounts of unlabeled real data for training, (2) ensure that model performance transfers between the two domains of data and simulation seamlessly, and (3) potentially build more generalizable, transferable, and powerful ML models that accelerate scientific discovery.

This framework, sometimes referred to as the foundation model (FM)~\cite{foundationModels} approach, has powered major advances in natural language processing, computer vision, audio/video processing, and learning across these different modalities of data.
These FMs, e.g., GPT-3~\cite{NEURIPS2020_1457c0d6}, are often large with up to $\mathcal{O}(10^9)$ parameters, trained on massive unlabeled data sets, serve as powerful backbones that extract key information from data, and can be adapted for different downstream tasks.

% related work and motivation
Related work includes using self-supervision for jet tagging~\cite{Dillon:2021gag}, resimulation based SSL~\cite{harris2024resimulationbased}, masked particle modeling~\cite{heinrich2024masked}, and generative pretraining~\cite{birk2024omnijetalpha}.
These studies have laid the groundwork for developing foundation models tailored to the unique challenges of LHC physics, highlighting the potential of various pretraining techniques. 
However, to fully harness the capabilities of a foundation model, it is crucial to demonstrate its ability to utilize the vast, mostly untapped amounts of unlabeled data produced by the LHC. 
Our study focuses on this critical aspect by progressively increasing the size of the pretraining dataset and assessing the resultant performance enhancements, thereby underscoring the model's scalability and effectiveness in leveraging large-scale unlabeled data. 
Our software is available at \url{https://github.com/JavierZhao/JetCLR/tree/JetClass}.

This paper is organized as follows.
Section~\ref{sec:dataset} describes the chosen datasets and how we use them to represent real data and simulation, respectively.
Section~\ref{sec:pretraining} describes the pretraining objective and training process we employ.
Section~\ref{sec:finetuning} discusses our finetuning stage.
The results are presented in Section~\ref{sec:results}.
Finally, Section~\ref{sec:summary} provides a summary and outlook.

\section{Datasets and experimental setup}
\label{sec:dataset}

\subsection{Datasets}
In this study, we used two datasets: JetClass~\cite{Qu2022,qu_2022_6619768}, a large dataset with 100 million jets that stands in for unlabeled data, and Top Tagging~\cite{Kasieczka:2021xcg,Kasieczka2019TopQuark}, a smaller dataset with 1.2 million jets that stands in for labeled simulation. 

The JetClass dataset contains ten distinct classes of jets, each corresponding to a unique particle decay channel, making it suitable to serve as a pretraining dataset.
The background jets are initiated by gluons and quarks ($\PQq/\Pg$).
The signal classes include five decay channels of the Higgs boson ($\PH\to\bbbar$, $\PH\to\ccbar$, $\PH\to\Pg\Pg$, $\PH\to 4\PQq$, and $\PH \to \ell \nu\PQq\PQq'$); two decay channels of the top quark ($\PQt \to \PQb\PQq\PQq'$ and $\PQt \to \PQb\ell\nu$); and the $\PW$ and $\PZ$ bosons decaying into a pair of quarks ($\PW \to \PQq\PQq'$ and $\PZ \to \qqbar$). 

The Top Tagging dataset only contains two classes of jets: top quark jets and QCD background jets, making it suitable for a downstream classification task.
\subsection{Experiment setup}
\label{subsec:Experiment setup}
For both the JetClass dataset and the Top Tagging dataset, we used six particle-level kinematic features as inputs, described in Table~\ref{table:kinematic features}. 
We use the ``raw'' $\eta$ and $\phi$ instead of the relative $\Delta \eta$ and $\Delta \phi$.
The reason is that as part of our contrastive learning approach, we augment the jets by translating the $\eta$ and $\phi$ coordinates of the individual particles.
The relative $\Delta \eta$ and $\Delta \phi$ would be invariant under this augmentation because all particles would have been shifted by the same amount.

\begin{table}[h!]
\centering
\begin{tabular}{ll}
Variable & Definition \\
\hline
$\eta$ & pseudorapidity \\
$\phi$ & azimuthal angle\\
$\log \pt$ & logarithm of the particle's transverse momentum $\pt$ \\
$\log E$ & logarithm of the particle’s energy \\
$\log(\pt/\pt^\mathrm{jet})$ & logarithm of the particle's $\pt$ relative to the jet $\pt$ \\
$\log(E/E^\mathrm{jet})$ & logarithm of the particle's energy relative to the jet energy
\end{tabular}
\caption{Particle input kinematic features used in training~\cite{Qu2022}}
\label{table:kinematic features}
\end{table}

\section{Self-supervised pretraining}
\label{sec:pretraining}
Our pretraining strategy draws on the innovative framework of JetCLR~\cite{Dillon:2021gag}, which introduces a contrastive learning paradigm specifically tailored for jet tagging.
Instead of using only 3 particle input features $(\pt, \eta, \phi)$ as in JetCLR, we took advantage of more features, as described in Section~\ref{subsec:Experiment setup}.

Central to the JetCLR approach, and by extension to our methodology, are the concepts of alignment and uniformity: bringing representations of similar jets---both augmented and original---closer together while distancing those of dissimilar jets.

\subsection{Augmentations}
\label{subsec:Augmentations}
\begin{figure}[htpb]
    \centering
    \includegraphics[width=1.0\textwidth]{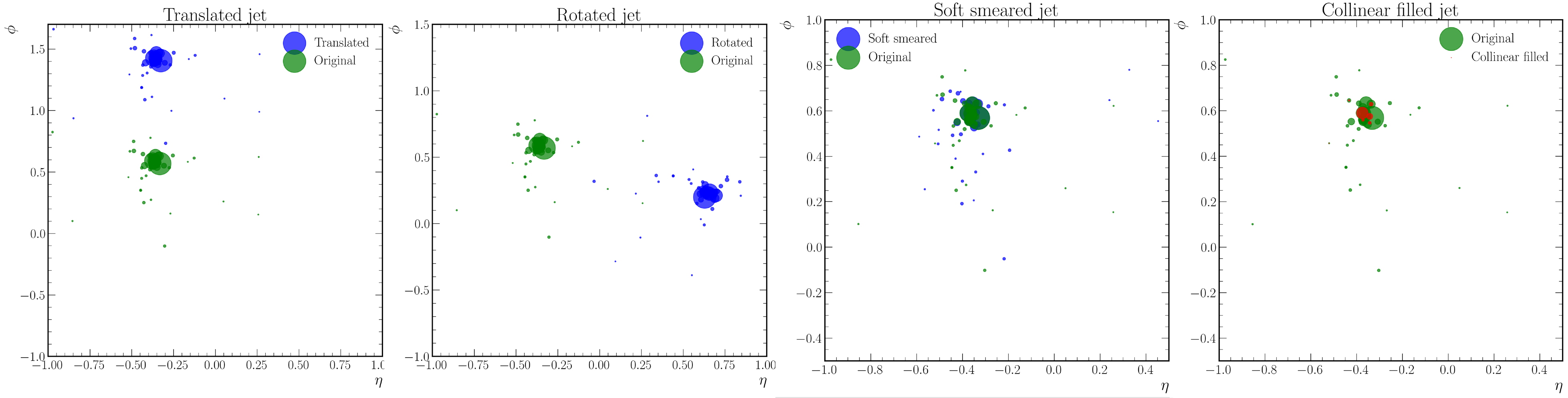}
    \caption{Illustration of the augmentations}
    \label{fig:Augmentations}
\end{figure}

To generate the positive pairs, we employed the same augmentation techniques as those detailed in the JetCLR framework~\cite{Dillon:2021gag}.
These augmentations are visualized in Fig.~\ref{fig:Augmentations}, which shows the angular distribution of jet particles, with the size of each marker being indicative of the particle's transverse momentum. 

\subsection{Training details}
\label{subsec:Training details}
To encode the input jets into representations, we used a transformer encoder~\cite{vaswani2023attention} with hyperparameters specified in Table ~\ref{table:hyperparameters}. 

In order to efficiently pretrain on a large number of jets, we made several optimizations to the existing code base from \cite{Dillon:2021gag} to speed up the pretraining.
First, we removed unnecessary CPU-GPU synchronizations, especially read-out from GPU for the purpose of recording losses.
In our optimized code base, this is done asynchronously to avoid synchronization barriers.
We also modified the default model dimensions to be multiples of 8 in order to make use of CUDA matrix multiplication kernels more efficiently.
In addition, we fused point-wise operations into a single CUDA kernel when computing the contrastive loss.
Last but not least, we utilized the Automatic Mixed Precision (AMP) package from PyTorch~\cite{NEURIPS2019_9015} to run certain parts of pretraining with lower precision, reducing memory usage and improving computational speed without significantly impacting the accuracy of results.
We used gradient clipping with a maximum norm of 0.1 and set the $\epsilon$ parameter to $10^{-4}$ in the Adam optimizer, in order to mitigate the numerical instability issues caused by the use of AMP.
These settings help control the scale of gradients, preventing them from becoming too large, which can lead to numerical instability during training.
These optimizations have improved our training speed significantly, by a factor of 3, compared with the original JetCLR implementation, with little impact on performance. 
\begin{table}[ht]
\centering
\begin{tabular}{l|c}
Hyperparameter             & Value \\ \hline
model (embedding) dimension          & 1024           \\
feed-forward hidden dimension        & 1024           \\
output dimension                     & 1024           \\
Self-attention heads              & 4              \\
Transformer layers            & 4              \\
Layers                            & 2              \\
dropout rate                         & 0.1            \\
\end{tabular}
\caption{Hyperparameters of the transformer encoder}
\label{table:hyperparameters}
\end{table}

\section{Supervised finetuning}
\label{sec:finetuning}
In the finetuning phase, we aim to preserve the rich, pretrained representations by appending only a single linear layer to the encoder, as opposed to a more elaborate multi-layer perceptron (MLP).
The rationale behind this decision was to prevent a sophisticated classifier from diluting the robust features that had been learned during the pretraining stage. 

During finetuning, we do not freeze the transformer encoder: both the encoder and the newly added linear layer are subject to training.
This allows for the fine adjustment of the features, ensuring that the encoder representations can adapt to the nuances of the task at hand while leveraging the generalizable knowledge acquired during pretraining.
This guides the model to maintain its powerful representational ability while also becoming more specialized for the targeted application.

\section{Results}
\label{sec:results}
\subsection{Transfer learning: From JetClass to Top Tagging}
\begin{figure}
    \centering
    \includegraphics[width=0.5\linewidth]{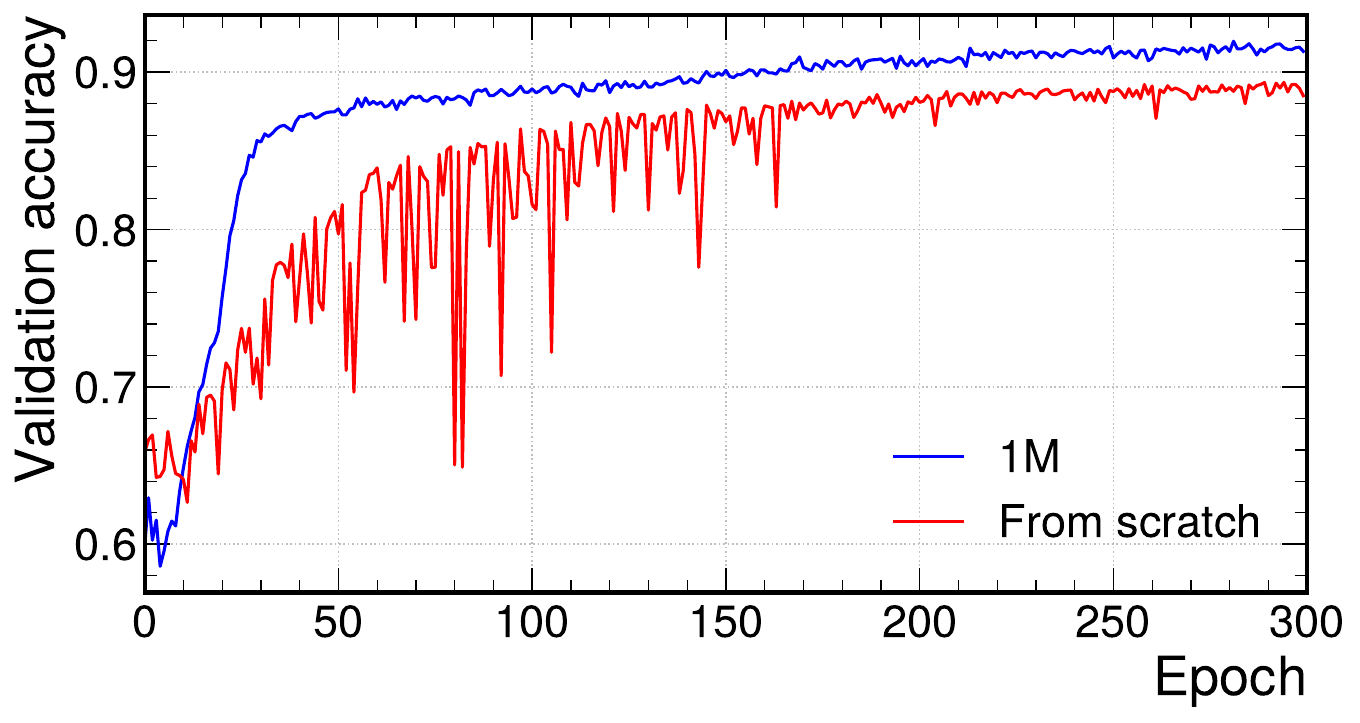}
    \caption{Comparison of the training performance of the pretrained model and the model trained from scratch}
    \label{fig:finetune_train}
\end{figure}

To demonstrate the generalizability and transferability of the features learned during the pretraining state, we pretrained a model on 1 million unlabeled jets from the JetClass dataset and then finetuned it on a certain number of labeled jets from the Top Tagging dataset. Fig ~\ref{fig:finetune_train} shows that despite the limited data, the pretrained model (blue curve) achieves higher accuracy and converges faster than the one trained from scratch (red curve). 

To further quantify the advantages of pretraining, we compared the  performances of the pretrained model against that of a counterpart initialized without pretraining.
Figures~\ref{fig:rej}~and~\ref{fig:acc} show the accuracy and rejection rates plotted against varying quantities of labeled training data, which show that the pretrained model consistently outperforms its from-scratch-trained counterpart. This is apparent not only in terms of accuracy but also in the efficiency of rejecting false positives.
As the volume of finetuning data increases, an improvement is noted for both models. However, the pretrained model achieves superior performance with a markedly smaller dataset—a clear testament to the pretraining phase's value. 

Additionally, the pretrained model demonstrates faster convergence, which is highlighted in Fig.~\ref{fig:epochs} comparing the number of epochs required to approach the final accuracy within a 1\% margin. 
Illustrated by the blue line in the graph, the pretrained model reaches the convergence threshold in fewer epochs over the entire span of labeled data presented, suggesting an optimized training that capitalizes on the learned features during pretraining to accelerate overall convergence.

\subsection{Demonstrating the power of dataset scaling through pretraining on different amounts of jets}
\begin{figure}[!ht]
    \centering
    \begin{minipage}{0.49\linewidth}
        \includegraphics[width=\linewidth]{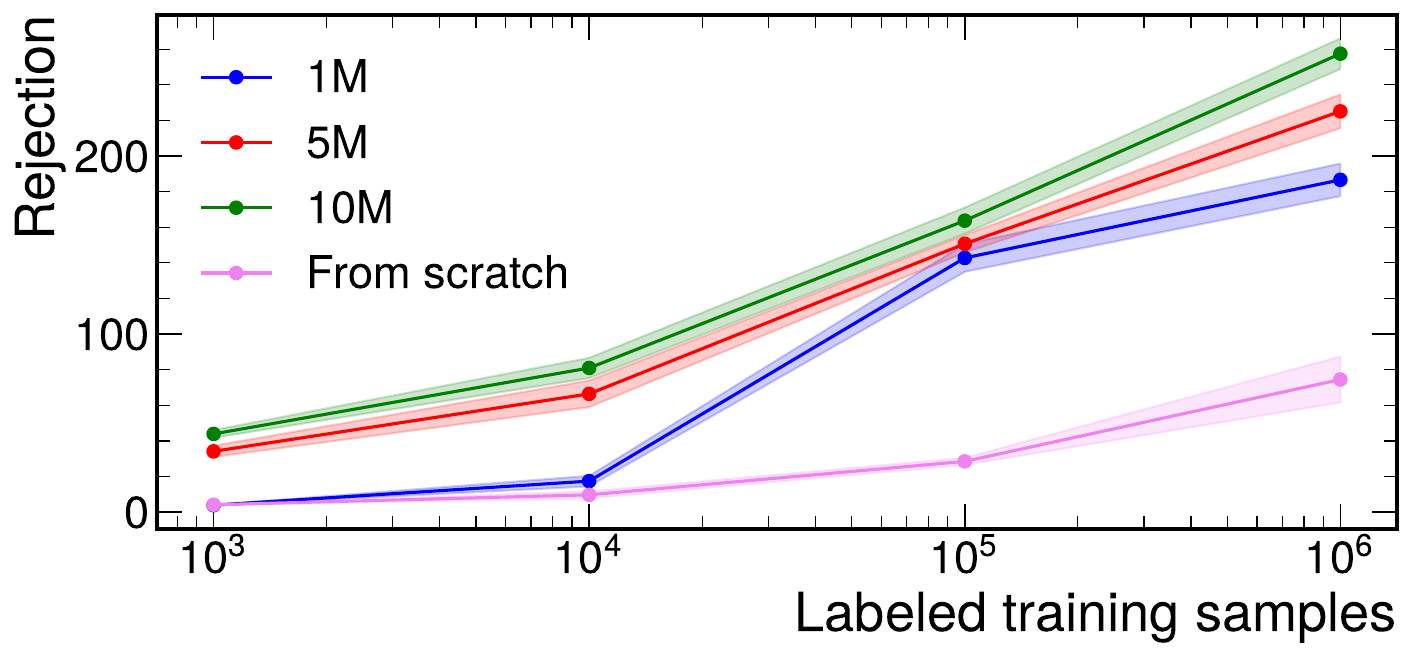}
        \caption{Background rejection at 50\% signal efficiency of different training strategies, with legend indicating the number of unlabeled training samples used in pretraining}
        \label{fig:rej}
    \end{minipage}
    \hfill
    \begin{minipage}{0.49\linewidth}
        \includegraphics[width=\linewidth]{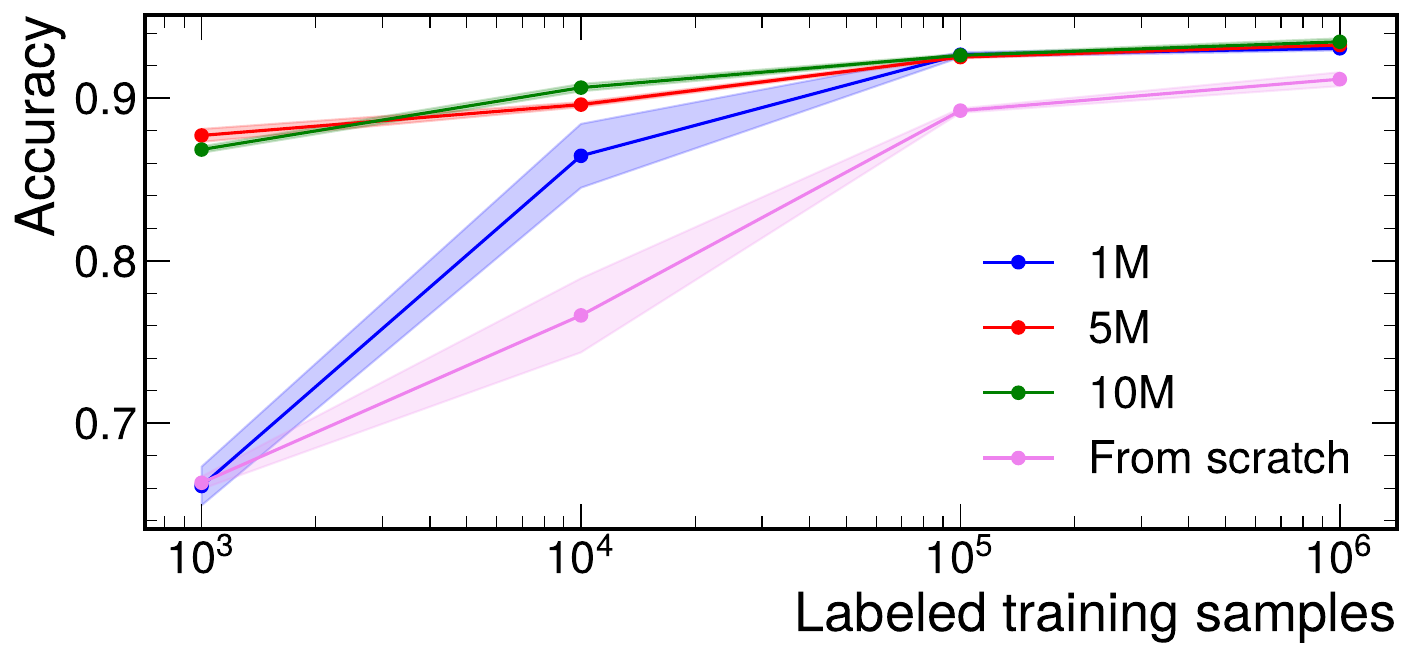}
        \caption{Accuracy of different training strategies, with legend indicating the number of unlabeled training samples used in pretraining}
        \label{fig:acc}
    \end{minipage}
\end{figure}

\begin{figure}[!ht]
    \centering
    \includegraphics[width=0.5\linewidth]{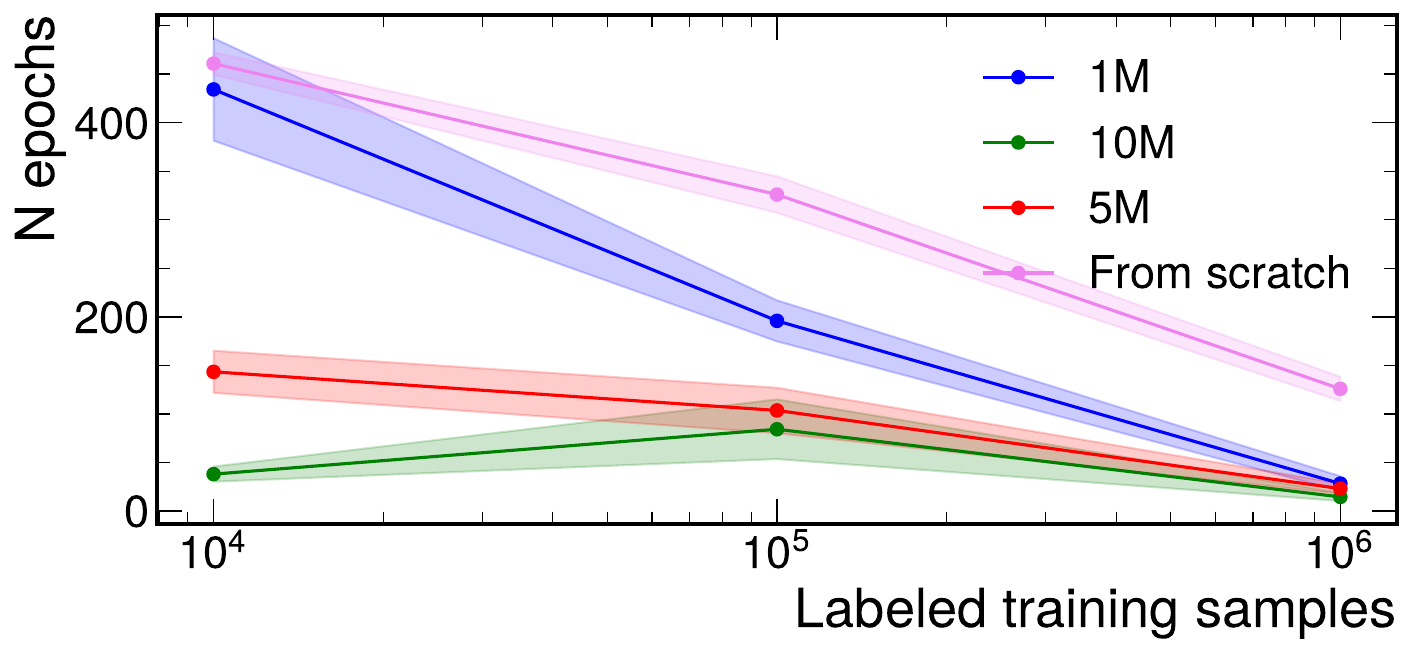}
    \caption{Number of epochs required to reach within 1\% of the final accuracy of different training strategies, with legend indicating the number of unlabeled training samples used in pretraining}
    \label{fig:epochs}
\end{figure}

To demonstrate the effectiveness of dataset scaling, we progressively increased the size of our pretraining dataset from 1 million jets to 5 million and finally to 10 million jets. We then analyzed the outcomes of models pretrained on these varying dataset sizes.
As illustrated in Figs.~\ref{fig:rej}~and~\ref{fig:acc}, the results show a clear trend: larger pretraining datasets significantly enhance model performance in terms of both rejection power and accuracy. Here, 'rejection power' refers to the model's ability to accurately reject background samples while correctly identifying signal samples.
In addition to enhancing model performance, larger pretraining datasets also allow the model to converge more quickly during finetuning.
As depicted in Fig.~\ref{fig:epochs}, as the sizes of the pretraining dataset grows, the number of epochs required to reach the final accuracy with a 1\% margin decreases across the range of labeled training samples used for finetuning.
These results demonstrate the feasibility of a foundation model that leverages the vast amounts of unlabeled data produced by the LHC.
This approach highlights the potential of using extensive, unlabeled datasets to enhance model capabilities significantly.
By training on increasingly larger datasets, we can tap into deeper insights and detect subtle patterns that are not apparent in smaller datasets. 

% \begin{figure}[!ht]
%     \centering
%     \includegraphics[width=0.7\linewidth]{figures/rej_pretrained_vs_from_scratch.pdf}
%     \caption{Background rejection at 50\% signal efficiency of different training strategies as a function of labeled training samples}
%     \label{fig:rej}
% \end{figure}

% \begin{figure}[!ht]
%     \centering
%     \includegraphics[width=0.7\linewidth]{figures/acc_pretrained_vs_from_scratch.pdf}
%     \caption{Accuracy of different training strategies as a function of labeled training samples}
%     \label{fig:acc}
% \end{figure}

\section{Summary and outlook}
\label{sec:summary}
In our study, we were able to show that 
\begin{enumerate}
    \item Through contrastive learning, a vanilla transformer encoder was able to learn useful representations of jets from unlabeled data.
    \item By pretraining on unlabeled data, the transformer encoder was able to learn the downstream task \textbf{faster} and with \textbf{fewer labeled training samples}, compared with one we trained from scratch.
    \item By scaling up the pretraining dataset, the model demonstrated enhanced performance and faster convergence, achieving higher accuracy and rejection rates with increased volumes of unlabeled data.
\end{enumerate}
Through large-scale pretraining followed by finetuning, our SSL approach has demonstrated \textbf{enhanced data efficiency}—requiring fewer labeled training samples to achieve superior performance compared to the fully supervised approach. Additionally, it offers \textbf{greater computational efficiency} by enabling the model to converge significantly faster than its fully supervised counterpart.
This paves the way for the use of unlabeled data in HEP and contributes to a better understanding the potential of SSL for scientific discovery.

\section{Acknowledgements}
\label{sec:Acknowledgements}
This work was supported by the Research Corporation for Science Advancement (RCSA) under grant \#CS-CSA-2023-109, Alfred P. Sloan Foundation under grant \#FG-2023-20452, U.S. Department of Energy (DOE), Office of Science, Office of High Energy Physics Early Career Research program under Award No. DE-SC0021187, the DOE, Office of Advanced Scientific Computing Research under Award No. DE-SC0021396 (FAIR4HEP), and the U.S. National Science Foundation (NSF) Harnessing the Data Revolution (HDR) Institute for Accelerating AI Algorithms for Data Driven Discovery (A3D3) under Cooperative Agreement OAC-2117997.
This work was performed using the Pacific Research Platform Nautilus HyperCluster supported by NSF awards CNS-1730158, ACI-1540112, ACI-1541349, OAC-1826967, the University of California Office of the President, and the University of California San Diego's California Institute for Telecommunications and Information Technology/Qualcomm Institute. 
Thanks to CENIC for the 100\,Gpbs networks.

\bibliographystyle{cms_unsrt}
\bibliography{bibliography}
\end{document}